\documentstyle[aps,epsf,preprint]{revtex}

\newcommand{\vare}{\varepsilon}

\newcommand{\rk}{{\rm k}}

\newcommand{\bfp}{{\bf p}}

\newcommand{\bfx}{{\bf x}}
\newcommand{\bfk}{{\bf k}}

\newcommand{\balpha}{{\mbox{\boldmath$\alpha$}}}

\newcommand{\intinf}{\int^{\infty}_{-\infty}}

\newcommand{\lbr}{\langle}
\newcommand{\rbr}{\rangle}

\begin{document}
%
%%%%%%%%%%%%%%%%%%%%%%%%%%%%%%%%%%%%%%%%%%%%%%%%%%%%%%%%%%%%%%%%%%%%%%%%%
%
%%%%%%%%%%%%%%%%%%%%%%%%%%%%%%%%%%%%%%%%%%%%%%%%%%%%%%%%%%%%%%%%%%%%%%%%%
%
\title{Loop-after-loop contribution to the second-order Lamb shift
in hydrogenlike low-$Z$ atoms
 }
\author{V.~A.~Yerokhin$\footnote{e-mail: yerokhin@fn.csa.ru}$}

\address{
 Department of Physics, St. Petersburg State University,
 Oulianovskaya 1, Petrodvorets, St. Petersburg 198904, Russia\\
and\\
 Institute for High Performance Computing and Data Bases, Fontanka
118, St. Petersburg 198005, Russia\\ }
\date{\today}
\maketitle

\begin{abstract}
We present a numerical evaluation of the loop-after-loop contribution to the
second-order self-energy for the ground state of hydrogenlike atoms with low nuclear
charge numbers $Z$. The calculation is carried out in the Fried-Yennie gauge and
without an expansion in $Z \alpha$. Our calculation confirms the results of
Mallampalli and Sapirstein  and disagrees with the calculation by Goidenko and
coworkers. A discrepancy between different calculations is investigated. An accurate
fitting of the numerical results provides a detailed comparison with analytic
calculations based on an expansion in the parameter $Z \alpha$. We confirm the
analytic results of order $\alpha^2 (Z\alpha)^5$ but disagree with Karshenboim's
calculation of the $\alpha^2 (Z \alpha)^6 \ln^3(Z \alpha)^{-2}$ contribution.
\end{abstract}

\section*{Introduction}

In the low-$Z$ region, calculations of radiative corrections in bound-state QED have
historically relied on a (semi-) analytic expansion in powers of the external binding
field $Z \alpha$. Calculations based on this perturbative approach have made an
enormous advance during the last 50 years and achieved an excellent agreement with
experiments (see, for example, a recent review \cite{LowZQED}). However, calculations
in higher orders in $Z \alpha$ become increasingly complex, as the number of terms in
each higher order increases rapidly. Beside this, it is difficult to estimate the
contribution of unevaluated higher-order terms. These are the reasons why the exact
numerical treatment of radiative corrections is highly appreciated even in the low-$Z$
region. It allows to test the reliability of methods based on an expansion in $Z
\alpha$ and can provide even more accurate results than analytic perturbative
calculations. Some examples of this are the calculation of the self-energy correction
to the hyperfine splitting in muonium performed by Blundell and coworkers
\cite{Blundell97}, the calculation of the relativistic recoil correction for hydrogen
by Shabaev {\it et al.} \cite{Shabaev98}, and the evaluation of the first-order
self-energy correction for $Z=1-5$ by Jentschura {\it et al.} \cite{Jentschura99}.

The aim of the present work is a numerical evaluation of the loop-after-loop
contribution to the second-order Lamb shift of the ground state in hydrogen-like atoms
to all orders in $Z \alpha$ in the low-$Z$ region. Analytic calculations of the $Z
\alpha$-expansion coefficients for this contribution were previously performed by
Eides and coworkers \cite{Eides93}  and Pachucki \cite{Pachucki94} in order $\alpha^2
(Z \alpha)^5$ and by Karshenboim \cite{Karshenboim93} in order $\alpha^2 (Z \alpha)^6
\ln^3(Z \alpha)^{-2}$. The first calculation of the loop-after-loop correction without
an expansion in $Z \alpha$ was carried out by Mitrushenkov {\it et al.}
\cite{Mitrushenkov} for high-$Z$ atoms. Recently, this correction was calculated to
all orders in $Z \alpha$ for the entire range of nuclear charge numbers by Mallampalli
and Sapirstein \cite{Mallampalli98}. A fit to the data from Ref. \cite{Mallampalli98}
confirms the analytic result of order $\alpha^2 (Z \alpha)^5$ but it is in a
significant disagreement with Karshenboim's result of order $\alpha^2 (Z \alpha)^6
\ln^3(Z \alpha)^{-2}$. The subsequent calculation by Goidenko {\it et al.}
\cite{Goidenko99}, also non-perturbative in $Z \alpha$, shows to be compatible with
the analytic calculations. In this work, we perform an independent calculation of the
loop-after-loop correction and investigate possible reasons for the discrepancy
between different calculations. Relativistic units are used in this article
($\hbar=c=m=1$).

%
%%%%%%%%%%%%%%%%%%%%%%%%%%%%%%%%%%%%%%%%%%%%%%%%%%%%
%
\section{Basic formalism and numerical procedure}

The expression for the irreducible contribution of Fig 1a (we refer to it as the {\it
loop-after-loop} correction) reads
\begin{equation}\label{lal}
  \Delta E_{\rm lal} = \sum_{\vare_n\ne \vare_a} \frac{\lbr a| \Sigma_R(\vare_a)|n\rbr
    \lbr n| \Sigma_R(\vare_a)|a\rbr}{\vare_a-\vare_n} \ ,
\end{equation}
where $\Sigma_R$ denotes the renormalized self-energy operator, $|a\rbr$ indicates the
initial state and the summation is performed over the spectrum of the Dirac equation.
The term with $\vare_n=\vare_a$ corresponds to the reducible contribution and should
be calculated together with the remaining diagrams in Fig. 1. The self-energy operator
is defined by its matrix elements
\begin{eqnarray} \label{se}
\lbr a| \Sigma_R(\vare)|b \rbr &=& i\alpha \intinf d\omega
    \int d^3\bfx_1 d^3 \bfx_2 \psi^{\dag}_a(\bfx_1)
                            \nonumber \\
&& \times \alpha_{\mu} G(\vare-\omega,\bfx_1,\bfx_2) \alpha_{\nu}
        \psi_b(\bfx_2) D^{\mu \nu}(\omega, \bfx_{12})
                            \nonumber \\
&& - \delta m \int d^3\bfx \psi^{\dag}_a(\bfx)\beta
        \psi_b(\bfx) \ ,
\end{eqnarray}
where $\alpha_{\mu} = (1,\balpha)$; $\beta$, $\balpha$  are the Dirac matrices,
$G(\omega) = 1/(\omega-{\cal H}(1-i0))$ is the Dirac-Coulomb Green function, ${\cal
H}=(\balpha \cdot \bfp) + \beta m + V(\bfx)$ is the Dirac-Coulomb Hamiltonian, $\delta
m$ is the mass counterterm, and $D^{\mu \nu}(\omega, \bfx_{12})$ is the photon
propagator in a general covariant gauge
\begin{equation}\label{}
  D^{\mu \nu}(\omega, \bfx_{12}) =  \int \frac{d\bfk}{(2\pi)^3}
  e^{i\bfk \bfx_{12}} \left(-\frac{g^{\mu \nu}}{\rk^2+i0}+
 \left.  (1-\lambda) \frac{\rk^{\mu}\rk^{\nu}}{(\rk^2+i0)^2} \right)
\right|_{\rk^0=\omega} \ .
\end{equation}
To our knowledge, up to now all the practical self-energy calculations without an
expansion in $Z \alpha$ were carried out in the Feynman gauge ($\lambda = 1$) which is
technically the easiest choice of the gauge. While the usage of the Feynman gauge in
calculations of the self-energy matrix elements is natural in the high-$Z$ region, for
low $Z$ it is known to provide a spurious contribution of order $Z \alpha$ which
should be cancelled numerically to give a residual of order $(Z \alpha)^4$. This
spurious term is known to vanish in the Fried-Yennie gauge \cite{Abrikosov,FY}
($\lambda = 3$), which possesses remarkable infrared properties. Since the present
work is aimed to a calculation of the loop-after-loop correction in the low-$Z$
region, we use the fact that this contribution is invariant in any covariant gauge and
perform our calculations in the Fried-Yennie gauge.

A general method which was used here for the calculation of the self-energy matrix
elements can be found in Ref. \cite{Yerokhin99}, with some modifications due to a
non-diagonal nature of the matrix elements and the different gauge. The self-energy
matrix element is considered as a sum of two contributions originating from an
expansion of the bound electron propagator in terms of interactions with the external
field of the nucleus
\begin{equation}\label{se1}
\lbr a|\Sigma_R(\vare)|b \rbr = \lbr a|\Sigma^{(0+1)}_R(\vare)|b\rbr + \lbr a|
\Sigma^{(2+)}(\vare)|b \rbr \ .
\end{equation}
Here, the first term contains zero and one Coulomb interaction with the nucleus, and
the second term contains two and more interactions. They are calculated in momentum
and coordinate space, respectively.

The expression (\ref{lal}) for the loop-after-loop contribution contains a summation
of non-diagonal self-energy matrix elements over the whole spectrum of the Dirac
equation. To perform the summation, we use the B-splines method for the Dirac equation
developed by Johnson {\it et al.} \cite{Johnson88}. In this method, the infinite
summation in the spectral representation of the Green function with a fixed angular
momentum quantum number is replaced by a finite sum over basis-set functions. A
straightforward evaluation of the sum in Eq.~(\ref{lal}) implies a computation of many
self-energy matrix elements with highly-oscillating wave functions and is
computationally intensive. To reduce the computational time significantly, we define a
self-energy correction to the wave function, as proposed in Ref. \cite{Mitrushenkov}
\begin{equation}\label{se2}
  |\varphi_{SE}\rbr \equiv \Sigma_R(\vare_a)|a\rbr \ .
\end{equation}
According to Eqs.  (\ref{se1}) and (\ref{se2}), we write Eq.~(\ref{lal}) as
\begin{eqnarray}\label{se3}
  \Delta E_{\rm lal} &=& \int  \frac{d^3 \bfp_1}{(2\pi)^3} \frac{d^3\bfp_2}{(2\pi)^3}
  {\varphi_{SE}^{(0+1)}}^{\dag}(\bfp_1)
  G^{\rm red}(\vare_a,\bfp_1,\bfp_2) \varphi_{SE}^{(0+1)} (\bfp_2)
            \nonumber \\
&&   + 2 \int \frac{d^3 \bfp_1}{(2\pi)^3} d^3 \bfx_2
    {\varphi_{SE}^{(0+1)}}^{\dag}(\bfp_1)
  G^{\rm red}(\vare_a,\bfp_1,\bfx_2) \varphi_{SE}^{(2+)}(\bfx_2)
           \nonumber \\
&& + \int d^3 \bfx_1 d^3 \bfx_2 {\varphi_{SE}^{(2+)}}^{\dag}(\bfx_1)
  G^{\rm red}(\vare_a,\bfx_1,\bfx_2) \varphi_{SE}^{(2+)}(\bfx_2)  \ ,
\end{eqnarray}
where $G^{\rm red}(\vare_a,\bfx_1,\bfx_2)$, $G^{\rm red}(\vare_a,\bfp_1,\bfp_2)$, and
$G^{\rm red}(\vare_a,\bfp_1,\bfx_2)$ are the reducible Dirac-Coulomb Green functions
in coordinate, momentum, and mixed representations, respectively (by the {\it mixed}
representation we mean the Fourier transform over one coordinate variable).

As the first step of the numerical evaluation of Eq.~(\ref{se3}), the effective wave
functions $\varphi_{SE}^{(0+1)}(\bfp)$ and $\varphi_{SE}^{(2+)}(\bfx)$ are calculated
on a grid and stored in an external file. Their computation is not much more intensive
than an evaluation of a single self-energy matrix element. The most difficult part of
the calculation is the evaluation of $\varphi_{SE}^{(2+)}(\bfx)$. Working in the
Fried-Yennie gauge, we do not encounter severe cancellations between zero-, one-, and
many-potential terms, as occur in the case of the Feynman gauge. Still, significant
cancellations arise in the computation of the Green function $G^{(2+)}$ which contains
two and more interactions with the external field. In our implementation it is
evaluated by a point-by-point subtraction of the two first terms of the Taylor
expansion from the Dirac-Coulomb Green function (see Ref. \cite{Yerokhin99} for
details)
\begin{equation} \label{se4}
G^{(2+)}(\vare,x_1,x_2) = G(\vare,x_1,x_2)- \left. G(\vare,x_1,x_2) \right|_{Z=0}- Z
\left( \left. \frac{d}{dZ} G(\vare,x_1,x_2) \right|_{Z=0} \right) \ .
\end{equation}
To control the cancellations which arise in the low-$Z$ region, we monitor the
corresponding Wronskian difference $\Delta_{\kappa}^{(2+)}(\vare)$ which can be
calculated analytically ($\Delta_{\kappa}(\vare)$ is the Wronskian of the solutions of
the radial Dirac equation). Another numerical problem is the partial wave expansion.
Its convergence is somewhat slower in the case of the Fried-Yennie gauge than in the
Feynman gauge. In actual calculations we extended the summation up to sixty partial
waves. It was performed before all numerical integrations were carried out. The
remainder after the truncation of the sum was estimated taking into account the
asymptotic behaviour of the expansion terms. Several checks were made of calculations
of $\varphi_{SE}^{(0+1)}(\bfp)$ and $\varphi_{SE}^{(2+)}(\bfx)$. In one, we compared
the diagonal self-energy matrix elements to the known results for the first-order
self-energy contribution \cite{Mohr92,Jentschura99}. We also calculated the
irreducible contribution to the self-energy correction to the hyperfine splitting in
H-like atoms and found a good agreement with Ref. \cite{Blundell97}.

In the next step, we perform the radial integrations in Eq.~(\ref{se3}). The
Dirac-Coulomb Green function in coordinate space is evaluated using a finite basis set
constructed from B-splines, after a transformation to a piecewise-polynomial
representation as described in the Appendix. The momentum and the mixed
representations of the Green function are obtained by the direct numerical Fourier
transformation of the polynomial basis. After that, two-dimensional radial integrals
in Eq.~(\ref{se3}) are expressed as a linear combination of one-dimensional integrals
and can be easily evaluated up to a desirable precision. In actual calculations we
used a basis set consisting of 70 positive and 70 negative energy states. The
stability of the final results with respect to the size of the cavity and the number
of energy states was checked.

%
%%%%%%%%%%%%%%%%%%%%%%%%%%%%%%%%%%%%%%%%%%%%%%%%%%%%%%%%%%%%%
%
\section{Numerical results and discussion}

In Table I and Fig. 2 we present the results of our calculation of the loop-after-loop
contribution to the second-order Lamb shift of the ground state of hydrogenlike atoms,
expressed in the standard form
\begin{equation}\label{se5}
  \Delta E_{\rm lal} = \left( \frac{\alpha}{\pi}\right)^2 \frac{(Z \alpha)^5}
    {n^3} G_{\rm lal}(Z \alpha) \ .
\end{equation}
The results of two previous non-perturbative calculations of this correction are
presented in Table I and Fig. 2 as well. A comparison exhibits a good agreement of the
present calculation with the results of Mallampalli and Sapirstein
\cite{Mallampalli98} and a strong deviation from the results of Goidenko {\it et al.}
\cite{Goidenko99}.

Let us consider possible reasons for this discrepancy. The method used in Ref.
\cite{Goidenko99} is based on the multiple commutator approach combined with the
partial-wave renormalization (PWR) procedure. In the PWR method, the truncation of the
partial-wave expansion fulfils the role of the regularization parameter. This shows
that this method is non-covariant. Still, it can be used for the calculation of the
diagonal first-order self-energy matrix elements, for which the PWR procedure is known
to provide the correct result \cite{Persson93,Quiney93}. In Ref. \cite{Persson98} this
renormalization procedure was investigated for the self-energy correction to an
additional Coulomb screening potential. It was shown analytically that some spurious
terms arise in different parts of the total self-energy correction due to the
non-covariant nature of the renormalization procedure. According to Ref.
\cite{Persson98}, the spurious terms cancel each other if the perturbation is the
Coulomb potential. The cancellation of the spurious contributions in the total
self-energy correction holds no longer if the perturbation contains a magnetic photon
(see Ref. \cite{Yerokhin97} and a conclusion remark in Ref. \cite{Persson98}).

To consider this topic in more detail, we calculate the self-energy correction in the
presence of the perturbing potential $-\alpha/r$ both in the PWR scheme and using a
covariant renormalization. For this choice of a perturbing potential, the total
self-energy correction for a state $|a\rbr$ is $d/(dZ) \lbr a|
\Sigma_{R}(\vare_a)|a\rbr$. The results of calculations are listed in Table II. Our
calculation confirms the conclusions from Ref. \cite{Persson98} about $a)$ the
presence of spurious terms in different parts of the correction and $b)$ their
cancellation in the sum for this particular choice of a perturbing potential.
Summarizing, we conclude that it is possible that the PWR method applied to the
irreducible part of the second-order self-energy correction, can provide a nonzero
spurious contribution.

In order to compare our results with calculations based on an expansion in $Z \alpha$,
we approximate our data for the function $G_{\rm lal}$ by a least-squares fit with
five parameters $a_{50}$, $a_{63}$, $a_{62}$, $a_{61}$, and $a_{60}$ (the first index
of the $a$ coefficients indicates the power of $Z \alpha$, the second corresponds to
the power of $\ln (Z \alpha)^{-2}$). A fit to our numerical results in Table I yields
\begin{equation}\label{se7}
 a_{50} = 2.33 \mbox{\hspace*{1cm}} a_{63} = -1.1 \ .
\end{equation}
This is in a good agreement with the fitting coefficients from Ref.
\cite{Mallampalli98} ($a_{50} = 2.3$ or $2.8$ for different sets of data,
$a_{63}=-0.9$) but disagrees significantly with Karshenboim's analytic result
$a_{63}=-8/27$ \cite{Karshenboim93}.

In order to investigate this discrepancy in more detail, we note that the $Z
\alpha$-expansion calculations of the loop-after-loop correction in Refs.
\cite{Eides93,Pachucki94,Karshenboim93}  were performed in the Fried-Yennie gauge like
in the present work and, therefore, it is possible to compare the calculations on
intermediate stages. So, we expand the inner electron propagators in diagram Fig. 1a
in terms of interactions with the nuclear binding potential and calculate the first
six terms of the expansion separately. The corresponding Feynman diagrams are
presented in Fig. 3. These diagrams do not contain $\varphi_{SE}^{(2+)}(\bfx)$ which
is the most difficult part of the calculation. Therefore, we were able to calculate
them for very low fractional $Z$. This is important for a reliable fitting of our data
which vary very fast in the vicinity of $Z=0$. The remainder behaves more smoothly in
the low-$Z$ region and its fitting is easier. In the calculation of the diagrams shown
in Fig. 3, we use closed analytical expressions for the Dirac Green function with zero
and one Coulomb interaction. In this way we eliminate the numerical uncertainty due to
the finite basis set representation of the Green function. The numerical results for
each diagram in Fig. 3 were approximated by least-squares fits with eight or seven
parameters $a_{50}$, $a_{6i}$ ($i=3,\ldots,0$), $a_{7i}$ ($i=3,2,1$) (in the last case
$a_{71}$ was omitted). In order to reduce the statistical uncertainty of the fitting
procedure, a large number of points (twenty or more) was used. The stability of the
fitting coefficients was checked with respect to the number of points, minimal and
maximal nuclear charge numbers, and different fits. The numerical results and the
fitting coefficients for diagrams in Fig. 3 are listed in Table III.

We found a good agreement with results from Refs. \cite{Eides93,Pachucki94} for the
coefficient $a_{50}$ and with Ref. \cite{Karshenboim93} for the coefficient $a_{63}$
originating from diagram Fig. 3f. The only discrepancy with the analytical
calculations originates from diagram Fig. 3c. While this diagram should not contribute
to order $\alpha^2 (Z \alpha)^6 \ln^3(Z \alpha)^{-2}$ according to Karshenboim, our
calculation shows the presence of a cubed logarithm with coefficient $a_{63} =
-0.652(30)$.

Summarizing, we conclude that our calculation of the loop-after-loop correction
confirms the analytic result of Refs. \cite{Eides93,Pachucki94} for the coefficient
$a_{50}$ ($a_{50} = 2.3$). A fit to the numerical results yields
\begin{equation}\label{se8}
 a_{63} = -0.958(30) \mbox{\hspace*{1cm}} a_{62} = 3.3(5)
\end{equation}
for the diagrams shown in Fig. 3, and
\begin{equation}\label{se9}
 a_{63} = -0.05(7) \mbox{\hspace*{1cm}} a_{62} = 1.2(8)
\end{equation}
for the non-perturbative remainder.

We note a remarkably slow convergence of the $Z\alpha$-expansion for the
loop-after-loop contribution to the second-order Lamb shift. As an illustration, in
Fig. 4 we plot the contributions of the first one, two, and three expansion terms
together with the non-perturbative results. The expansion coefficients are taken from
Eqs.~(\ref{se8}) and (\ref{se9}). One can see that even for hydrogen the contribution
of the first three expansion terms covers only about 50\% of the total result. To
obtain a reasonable fit to the numerical data even for very low $Z$, it is necessary
to take into account at least four first expansion terms. This fact shows the
necessity for non-perturbative (in $Z\alpha$) calculations of the total second-order
Lamb shift in the low-$Z$ region.

\section*{Acknowledgments}
I would like to thank Prof. Shabaev for his guidance and continued interest during the
course of this work. Valuable conversations with S. G. Karshenboim and T. Beier are
acknowledged. I am grateful for the kind hospitality to Prof. Eichler during my visit
in autumn 1999. This work was supported by the Russian Foundation for Basic Research
(Grant No.~98-02-18350) and by the program "Russian Universities. Basic Research"
(project No. 3930).

%%%%%%%%%%%%%%%%%%%%%%%%%%%%%%%%%%%%%%%%%%%%%%%%%%%%%%%%%%%%%%%%%%%%%%%%%%%
%
%       Table 1
%
%%%%%%%%%%%%%%%%%%
\begin{table}
 \caption{The loop-after-loop contribution
to the second-order Lamb shift of the ground state of hydrogenlike atoms expressed in
terms of the function $G_{\rm lal}(Z \alpha)$ defined by Eq.~(\ref{se5}).}
\vspace*{0.5cm}
\begin{tabular}{dddd}
\hline
$Z$ & This work & Ref.\cite{Mallampalli98} & Ref.\cite{Goidenko99}\\ \hline
0.5 & $-$1.56(7)   &  $-$1.5(1)    &       \\
0.8 & $-$2.36(5)   &             &       \\
1   & $-$2.75(4)   &  $-$2.87(5)   &       \\
1.5 & $-$3.449(9)  &  $-$3.47(2)   &       \\
2   & $-$3.919(7)  &  $-$3.965(15) &       \\
3   & $-$4.476(3)  &  $-$4.50(1)   &  $-$2.101 \\
4   & $-$4.772(3)  &  $-$4.77(1)   &  $-$2.311 \\
5   & $-$4.927(2)  &  $-$4.931(5)  &  $-$2.485 \\
6   & $-$4.997(1)  &             &  $-$2.599 \\
7   & $-$5.015(1)  &  $-$5.016(3)  &  $-$2.694 \\
8   & $-$4.998(1)  &             &  $-$2.659 \\
9   & $-$4.958(1)  &             &  $-$2.642 \\
10  & $-$4.902(1)  &  $-$4.9016(14)&  $-$2.601 \\
12  & $-$4.762(1)  &             &       \\
15  & $-$4.523(1)  &  $-$4.5218(6) &       \\
20  & $-$4.122(1)  &  $-$4.1217(3) &  $-$2.568 \\
\hline
\end{tabular}
\end{table}

%%%%%%%%%%%%%%%%%%%%%%%%%%%%%%%%%%%%%%%%%%%%%%%%%%%%%%%%%%%%%%%%%%%%%%%%%%%
%
%       Table 2
%
%%%%%%%%%%%%%%%%%%
\begin{table}
\caption{The self-energy correction in the presence of the perturbing potential
$-\alpha/r$ calculated both in the partial-wave renormalization scheme (PWR) and using
a covariant renormalization (CR). $\Delta E_{\rm ir}$ is the irreducible contribution
(known also as {\it perturbed orbital} contribution), $\Delta E_{\rm vr}$ denotes the
sum of the reducible and the vertex contribution. The results are compared with the
derivative of the first-order self-energy contribution $\Delta E_{SE}$ with respect to
the nuclear charge number $Z$. The calculation is performed in the Feynman gauge for a
point nucleus.} \vspace*{0.5cm}
\begin{tabular}{cccccccc}
\hline
 $Z$  & $\Delta E_{\rm ir}^{\rm PWR}$
              & $\Delta E_{\rm ir}^{\rm CR}$
                        & $\Delta E_{\rm vr}^{\rm PWR}$
                                   & $\Delta E_{\rm vr}^{\rm CR}$
                                             & $\Delta E_{\rm total}^{\rm PWR}$
                                                       & $\Delta E_{\rm total}^{\rm CR}$
                                                                 & $\Delta E_{\rm SE}/dZ$
        \\  \hline
 20 & 0.02003 & 0.00813 & $-$0.00899 & 0.00289 & 0.01104 & 0.01102 & 0.01102 \\
 30 & 0.03969 & 0.02137 & $-$0.01080 & 0.00750 & 0.02888 & 0.02886 & 0.02885 \\
 50 & 0.10722 & 0.07372 & $-$0.00892 & 0.02460 & 0.09830 & 0.09832 & 0.09832 \\
 70 & 0.23749 & 0.18248 &  0.00197 & 0.05698 & 0.23946 & 0.23946 & 0.23946 \\
 92 & 0.55983 & 0.46257 &  0.03473 & 0.13199 & 0.59456 & 0.59456 & 0.59456 \\
 \hline
\end{tabular}
\end{table}

%%%%%%%%%%%%%%%%%%%%%%%%%%%%%%%%%%%%%%%%%%%%%%%%%%%%%%%%%%%%%%%%%%%%%%%%%%%
%
%       Table 2
%
%%%%%%%%%%%%%%%%%%
\begin{table}
 \caption{The contributions of the diagrams shown in Fig. 3, expressed in
 terms of the function $G_{\rm lal}(Z \alpha)$ defined by Eq.~(\ref{se5}).
 The numerical results for the first three coefficients of the $Z \alpha$-expansion
 corresponding to two different fits are listed and compared to
 the analytical calculations.}
\vspace*{0.5cm}
\begin{tabular}{dddddddd}
\hline
 $Z$ &Fig. 3a  &Fig. 3b &Fig. 3c   &Fig. 3d  &Fig. 3e &Fig. 3f  &Fig. 3(a-f) \\ \hline
 0.1 &  0.0072 & 9.1740 &  $-$7.9975 & $-$0.0099 & 0.1654 & $-$0.7931 &  0.5460 \\
 0.2 &  0.0121 & 9.0885 &  $-$8.4197 & $-$0.0176 & 0.2709 & $-$1.1941 & $-$0.2599 \\
 0.4 &  0.0198 & 8.9417 &  $-$8.9281 & $-$0.0306 & 0.4334 & $-$1.7448 & $-$1.3086 \\
 0.7 &  0.0283 & 8.7535 &  $-$9.3648 & $-$0.0472 & 0.6204 & $-$2.3078 & $-$2.3176 \\
 1.0 &  0.0347 & 8.5886 &  $-$9.6271 & $-$0.0618 & 0.7705 & $-$2.7177 & $-$3.0128 \\
 1.5 &  0.0425 & 8.3471 &  $-$9.8763 & $-$0.0831 & 0.9730 & $-$3.2199 & $-$3.8168 \\
 2.0 &  0.0478 & 8.1349 &  $-$9.9989 & $-$0.1019 & 1.1372 & $-$3.5886 & $-$4.3696 \\
 3.0 &  0.0534 & 7.7705 & $-$10.0572 & $-$0.1346 & 1.3949 & $-$4.1007 & $-$5.0738 \\
 5.0 &  0.0526 & 7.1957 &  $-$9.8680 & $-$0.1876 & 1.7519 & $-$4.6721 & $-$5.7274 \\
 7.0 &  0.0425 & 6.7505 &  $-$9.5453 & $-$0.2309 & 1.9946 & $-$4.9565 & $-$5.9451 \\
10.0 &  0.0177 & 6.2353 &  $-$9.0170 & $-$0.2852 & 2.2458 & $-$5.1386 & $-$5.9419 \\
15.0 & $-$0.0376 & 5.6313 &  $-$8.2016 & $-$0.3600 & 2.5111 & $-$5.1632 & $-$5.6201 \\
20.0 & $-$0.1023 & 5.2261 &  $-$7.5211 & $-$0.4248 & 2.6849 & $-$5.0591 & $-$5.1963 \\ \hline
\multicolumn{8}{l}{
Analytic results \cite{Eides93,Pachucki94,Karshenboim93}: }              \\
$a_{50}$& 0    &  9.284 &   $-$6.984 &    0    &   0    &  0      &  2.300 \\
$a_{63}$& 0    &  0     &    0     &    0    &   0    & $-$0.296  & $-$0.296 \\ \hline
\multicolumn{8}{l}{
Eight-parameter fit: }                                               \\
$a_{50}$& 0.000& 9.284  &   $-$6.985 &  0.000  & 0.000  &  0.001  &  2.300 \\
$a_{63}$& 0.000& 0.001  &   $-$0.658 &  0.002  &-0.003  & $-$0.304  & $-$0.963 \\
$a_{62}$& 0.02 &-0.09   &    3.1   & $-$0.07   & 1.17   & $-$0.75   &  3.34  \\ \hline
\multicolumn{8}{l}{
Seven-parameter fit: }                                               \\
$a_{50}$& 0.000 &  9.285&  $-$6.987  &  0.000  & 0.000  &  0.000  &  2.298 \\
$a_{63}$&-0.001 & $-$0.003&  $-$0.646  &  0.002  &-0.005  & $-$0.301  & $-$0.952 \\
$a_{62}$&-0.01  & $-$0.01 &   2.82   & $-$0.08   & 1.22   & $-$0.81   &  3.12  \\ \hline
\end{tabular}
\end{table}

%
%%%%%%%%%%%%%%%%%%%%%%%%%%%%%%%%%%%%%%%%%%%%%%%%%%%%%%%%%%%%%
%
\appendix
\section*{Piecewise-polynomial representation of the Green function}

The B-splines method for the Dirac equation \cite{Johnson88} provides a finite set of
radial wave functions with a fixed angular momentum quantum number which can be
written in the form
\begin{equation}\label{ap1}
  \varphi_{\kappa,n}^i(x) = \frac1{x} \sum_k c^i_k(\kappa,n,l) \,
        (x-x_l)^k \ ,
\end{equation}
assuming that $x\in \left[ x_l,x_{l+1} \right]$. Here $x_l$ is the radial grid, the
index $i=1,2$ indicates the upper and the lower component of the radial wave function,
$n$ numbers the wave functions in the set, and $\kappa$ is the angular momentum
quantum number. The radial Green function, defined by
\begin{equation}\label{ap2}
  G^{ij}_{\kappa}(\vare,x_1,x_2) = \sum_n
  \frac{\varphi_{\kappa,n}^i(x_1)\varphi_{\kappa,n}^j(x_2)}{\vare-\vare_n} \ ,
\end{equation}
can be written in the piecewise-polynomial representation as
follows:
\begin{equation}\label{ap3}
  G_{\kappa}^{ij}(\vare,x_1,x_2) = \frac1{x_1x_2} \sum_{k_1k_2}
      A^{ij}_{k_1k_2}(\vare,\kappa,l_1,l_2) (x-x_{l_1})^{k_1} (x-x_{l_2})^{k_2}\ ,
\end{equation}
where $x_1\in \left[ x_{l_1},x_{l_1+1} \right]$, $x_2\in \left[
x_{l_2},x_{l_2+1} \right]$. The coefficients $A^{ij}_{k_1k_2}$
are
\begin{equation}\label{ap4}
    A^{ij}_{k_1k_2}(\vare,\kappa,l_1,l_2) =
    \sum_n \frac{c^i_{k_1}(\kappa,n,l_1) c^j_{k_2}(\kappa,n,l_2)}
        {\vare-\vare_n} \ .
\end{equation}

The radial Green function in momentum space can be written in the same way using
Fourier transformed basic polynomials
\begin{eqnarray}\label{ap5}
  G^{ij}_{\kappa}(\vare,p_1,p_2) &=&  \sum_{l_1l_2} \sum_{k_1k_2}
      A^{ij}_{k_1k_2}(\vare,\kappa,l_1,l_2)
      \Pi^{ik_1}_{l_1}(p_1) \Pi^{jk_2}_{l_2}(p_2) \
      , \\
   \Pi^{ik}_{l}(p) &=& 4\pi s(L_i) \int_{x_l}^{x_{l+1}} dx\,
       x(x-x_l)^k j_{L_i}(px) \ ,
\end{eqnarray}
where $L_{1,2} = |\kappa\pm 1/2|-1/2$; $s(L_1)=1$, $s(L_2) = -\kappa/|\kappa|$; and
$j_L(z)$ denotes the spherical Bessel function.

\newpage
%%%%%%%%%%%%%%%%%%%%%%%%%%%%%%%%%%%%%%%%%%%%%%%%%%%%%%%%%%%%%%%%%%%%%%%%%%%
%
%       Figure 1
%
%%%%%%%%%%%%%%%%%%
\begin{figure}
\centerline{ \mbox{ \epsfxsize=\textwidth \epsffile{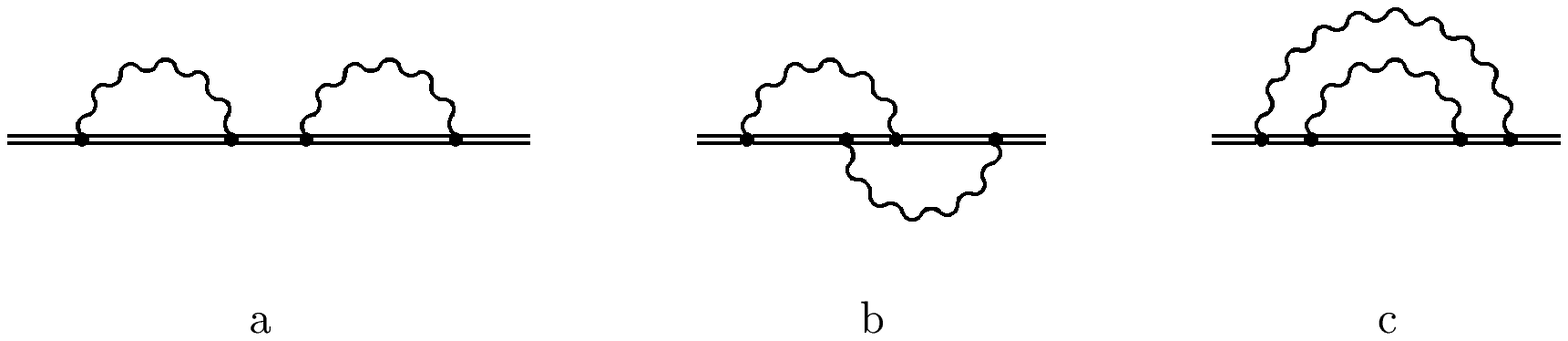} } }
\caption{One-electron self-energy Feynman diagrams of second order in $\alpha$.}
\end{figure}

\newpage
%%%%%%%%%%%%%%%%%%%%%%%%%%%%%%%%%%%%%%%%%%%%%%%%%%%%%%%%%%%%%%%%%%%%%%%%%%%
%
%       Figure 2
%
%%%%%%%%%%%%%%%%%%
\begin{figure}
\centerline{ \mbox{\epsfysize=\textwidth \epsffile{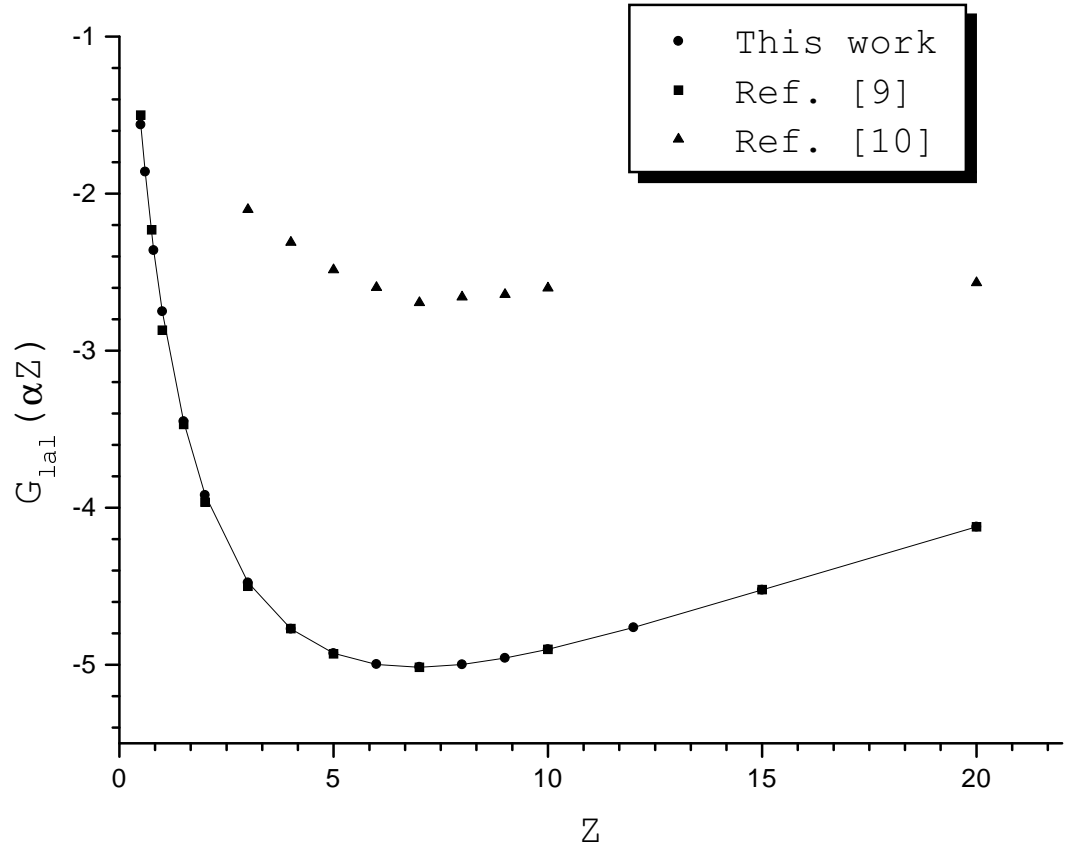} } } \vspace*{0.5cm}
\caption{The function $G_{\rm lal}(Z \alpha)$ in different calculations. The solid
line indicates a fit to our numerical results.}
\end{figure}

\newpage
%%%%%%%%%%%%%%%%%%%%%%%%%%%%%%%%%%%%%%%%%%%%%%%%%%%%%%%%%%%%%%%%%%%%%%%%%%%
%
%       Figure 3
%
%%%%%%%%%%%%%%%%%%
\begin{figure}
\centerline{ \mbox{ \epsfxsize=\textwidth \epsffile{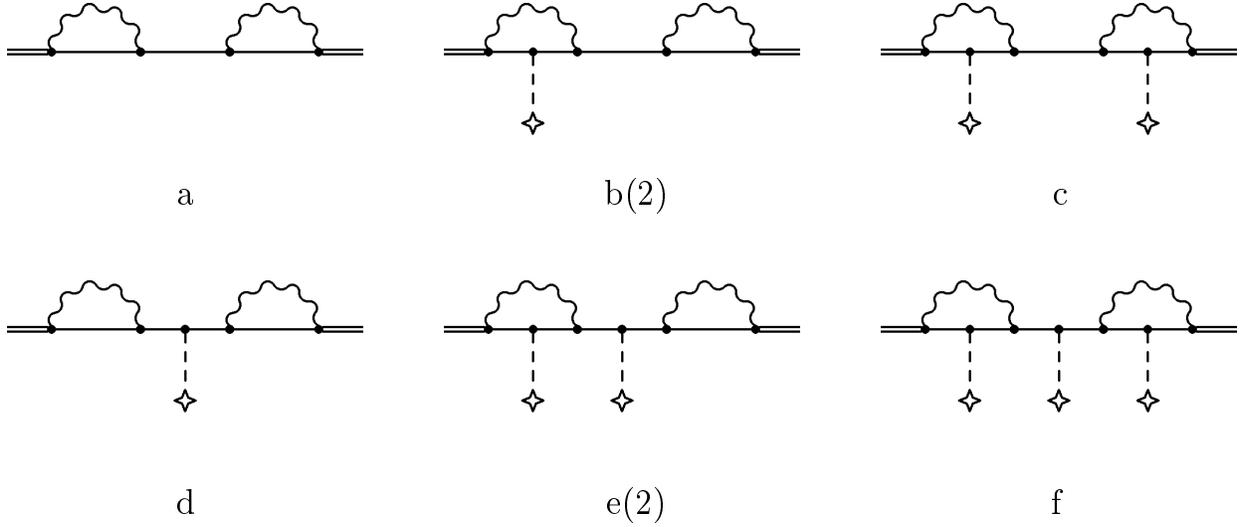} } } \vspace*{0.5cm}
\caption{Diagrams obtained from Fig. 1a by expansion of the inner electron propagators
in terms of interactions with the nuclear binding potential. A double line denotes the
electron in the field of the nucleus. A single line indicates the free electron. A
dashed line denotes a Coulomb interaction with the nucleus. Some diagrams are counted
twice, as is denoted by "(2)". }
\end{figure}

\newpage
%%%%%%%%%%%%%%%%%%%%%%%%%%%%%%%%%%%%%%%%%%%%%%%%%%%%%%%%%%%%%%%%%%%%%%%%%%%
%
%       Figure 4
%
%%%%%%%%%%%%%%%%%%
\begin{figure}
\centerline{
\mbox{
\epsfxsize=\textwidth \epsffile{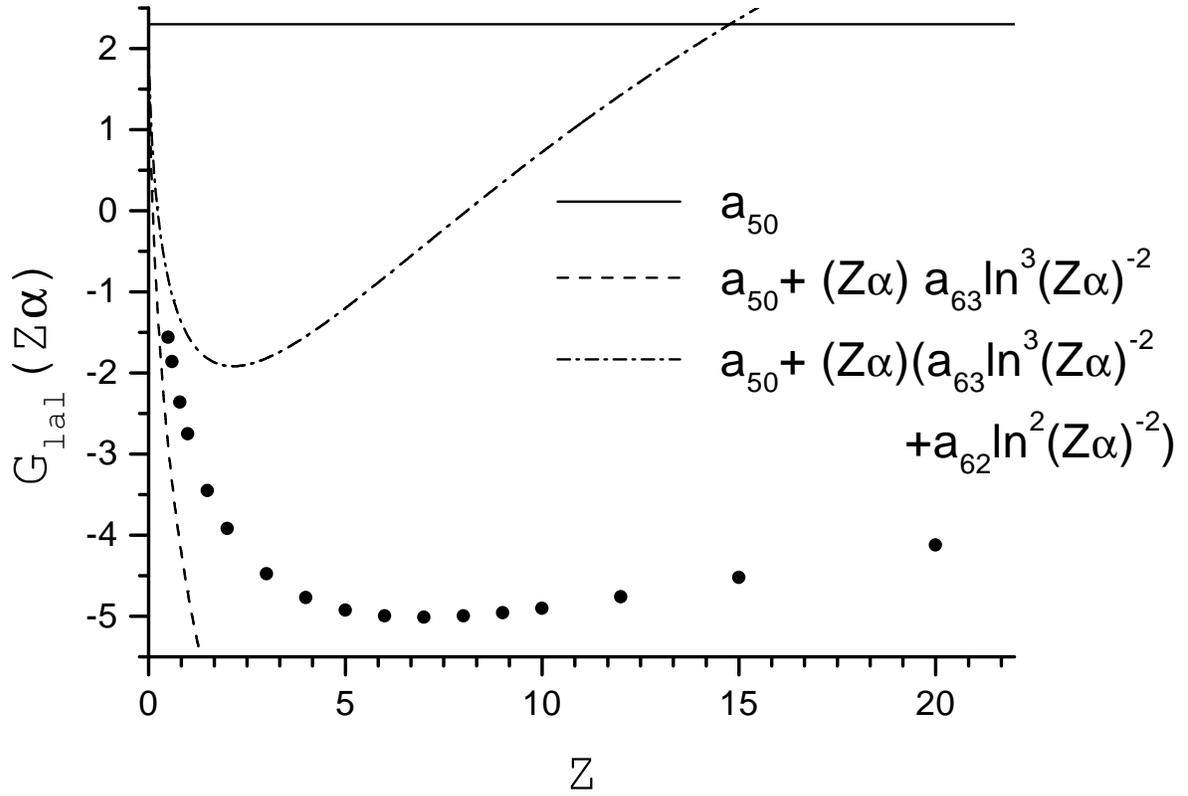}
}
}
\vspace*{0.5cm}

\caption{The non-perturbative (in $Z\alpha$) function $G_{\rm lal}(Z \alpha)$ and the
contributions of the first one, two, and three terms of its expansion in $Z\alpha$.
Dots indicate the non-perturbative results. The expansion coefficients are taken from
Eqs.~(\ref{se8}) and (\ref{se9}). }
\end{figure}

\end{document}